\documentclass[10pt,journal,compsoc]{IEEEtran}
\ifCLASSINFOpdf
\else
\fi
\usepackage{subcaption}
\usepackage{graphicx}
\usepackage[ruled, linesnumbered]{algorithm2e}
\usepackage{amsmath}
\usepackage{amsfonts}
\usepackage{multirow}
\usepackage{wasysym}
\usepackage{color} 
\usepackage{soul}
\usepackage{listings}
\usepackage{lipsum}
\usepackage{mathtools}
\usepackage{cuted}

\usepackage{hyperref} 
\usepackage{url}

\begin{document}
	%
	\title{A Self-adaptive Approach for Managing Applications and Harnessing Renewable Energy for Sustainable Cloud Computing}
	%
	%
	%
	
	\author{Minxian~Xu,~\IEEEmembership{}
		Adel N. Toosi,~\IEEEmembership{Member, ~IEEE,}
		and~Rajkumar Buyya,~\IEEEmembership{Fellow,~IEEE}
		\IEEEcompsocitemizethanks{\IEEEcompsocthanksitem M. Xu is with  Shenzhen Institutes of Advanced Technology, Chinese Academy of Sciences, China. Email: mx.xu@siat.ac.cn.
			\IEEEcompsocthanksitem A. N. Toosi is with Faculty of IT,
			Monash University, Australia. Email: adel.n.toosi@monash.edu.
			\IEEEcompsocthanksitem R. Buyya is with Cloud Computing and Distributed Systems (CLOUDS) lab, School of Computing and Information Systems, 
			University of Melbourne, Australia. Email: rbuyya@unimelb.edu.au.
		}
		
		\thanks{Manuscript received? ; revised }}
	
	%
	%
	\newcommand\MYhyperrefoptions{bookmarks=true,bookmarksnumbered=true,
		pdfpagemode={UseOutlines},plainpages=false,pdfpagelabels=true,
		colorlinks=true,linkcolor={black},citecolor={black},urlcolor={black},
		pdftitle={Bare Demo of IEEEtran.cls for Computer Society Journals},
		pdfsubject={Typesetting},
		pdfauthor={Michael D. Shell},
		pdfkeywords={Computer Society, IEEEtran, journal, LaTeX, paper,
			template}}

	\markboth{Journal of \LaTeX\ Class Files,~Vol.~6, No.~1, January~2007}%
	{Shell \MakeLowercase{\textit{et al.}}: Bare Demo of IEEEtran.cls for Journals}
	%




	\IEEEtitleabstractindextext{%
		\begin{abstract}
			Rapid adoption of Cloud computing for hosting services and its success is primarily attributed to its attractive features such as elasticity, availability and pay-as-you-go pricing model. However, the huge amount of energy consumed by cloud data centers makes it to be one of the 
			fastest growing sources of carbon emissions. Approaches for improving the energy efficiency include enhancing the resource utilization to reduce resource wastage and applying the renewable energy as the energy supply. This work aims to reduce the carbon footprint of the data centers by reducing the usage of brown energy and maximizing the usage of renewable energy.  Taking advantage of microservices and renewable energy, we propose a self-adaptive approach for the resource management of interactive workloads and batch workloads. To ensure the quality of service of workloads, a brownout-based algorithm for interactive workloads and a deferring algorithm for batch workloads are proposed.  We have implemented the proposed approach in a prototype system and evaluated it with web services under real traces. The results illustrate our approach can reduce the brown energy usage by 21\% and improve the renewable energy usage by 10\%.
			
		\end{abstract}
		
		
		}
	\maketitle
	
	\IEEEdisplaynontitleabstractindextext
	
	%
	\IEEEpeerreviewmaketitle

	\section{Introduction} \label{sec:introdution}
	Today's society and its organizations are becoming ever-increasingly dependent upon information and communication technologies (ICT) with software systems, especially web systems, largely hosted on cloud data centers. Clouds offer an exciting benefit to enterprises by removing the need for building own Information Technology (IT) infrastructures and shifting the focus from the IT and infrastructure issues to core business competence. Apart from the infrastructure, elasticity, availability, and pay-as-you-go pricing model are among many other reasons which led to the rise of cloud computing \cite{Kilcioglu2017}.  This massive growth in cloud solutions demanded the establishment of huge number of data centers around the world owned by enterprises and large cloud service providers such as Amazon, Microsoft, and Google to offer their services \cite{Chen2019WWW}.

	However, data centers hosting cloud services consume a large amount of electricity leading to high operational costs and high carbon footprint on the environment \cite{Jiang2019ISCA}. ICT sector nowadays consumes approximately 7\% of the global electricity, and it is predicted that the share will increase up to 13\% by 2030~\cite{Avgerinou2017}. Among this, the operation of data centers accounts for one of the fastest growing sources of carbon dioxide emissions~\cite{Whitehead2014}. In 2013, U.S. data centers solely consumed an estimated 91 billion kWh of electricity (equivalent to the two-year power consumption of all households in New York City) and this is projected to reach 140 billion kWh by 2020~\cite{Delforge2014}.

	One of the main sources of energy inefficiencies in data centers is \color{black} represented by servers,  \color{black}which are often utilized between 10\% to 50\% of their peak load~\cite{Pedram2012}. This issue is amplified by the fact that server machines in data centers do not exhibit complete energy proportionality, that is, servers do not consume electricity in proportion to their load~\cite{Beloglazov2015}. \color{black}Even though, cloud providers use techniques such as dynamic consolidation of virtual machines (VMs) ~\cite{AntonTPDS} to achieve energy saving and avoid underutilized servers, energy is still being wasted if cloud consumers hold many idle or under-utilized virtual machines up and running. \color{black}
 RightScale\footnote{https://www.rightscale.com/} states that the cloud consumers waste between 30-45\% of their total cloud consumption  \cite{Clement2017RightScale}

	In this regards, microservice architectures and technologies such as containers \cite{Newman2015}, that are steadily gaining adoption in industrial practice, provide a leap towards more efficient utilization of cloud resources. Containerization allows for higher resource utilization and reduction of cost by running multiple services on the same VM and providing a fine-grained \color{black}control \color{black}on resources. Traditional web applications are considered to be migrated from monolithic structure to microservices architecture \cite{Ren2018}.
	In this paper, we take advantage of container technology to reduce the energy consumption of the system.

		\color{black}Brownout \cite{XuBrownoutSurvey} is a self-adaptive approach to manage resources and applications in cloud computing systems. With brownout, the optional parts of applications can be dynamically deactivated/activated according to the system status. A control knob, called dimmer, is applied to represent the degree that brownout should be performed on the application's optional parts. Brownout can also be utilized for the management of microservices to improve resource usage in data centers. However, generally, there are always trade-offs that should be balanced, e.g. balancing energy consumption and system performance. \color{black}	
	
	Apart from self-contained microservices, renewable energy is another solution gaining momentum to address energy consumption concerns (i.e., the carbon footprint) of cloud computing.  In response to the climate change concerns and economic stimulus, many research initiatives have been launched to promote renewable energy use to power cloud data centers in recent years~\cite{Liu2012}\cite{Han2016}\cite{goiri2013parasol}. Many cloud providers also work on this goal by generating their own renewable energy or drawing power from a nearby renewable power plant.  For example, in January 2018, AWS achieved 50\% renewable energy usage by investing in clean energy activities including a commercial-scale wind farm in North Carolina.\footnote{ https://aws.amazon.com/about-aws/sustainability/}

	Renewable energy systems are shown to be extremely effective in reducing dependence on finite fossil fuels and decreasing environmental impacts. \color{black}Currently, all modern inverters have interfaces to select the source of power for either grid or batteries. However, power generation using photovoltaic (PV) solar energy can only be done during the daytime and the amount of \color{black}produced power depends on the weather and geographical location of the data center. A large solar power system with a sufficiently large battery setup to fully support workload is not economical. Therefore, we are looking into approaches to match energy consumption with the availability of renewable energy.
	In these approaches, cloud resource management systems need to support methods that allocate resources and schedule applications execution by preferring to finish them during the time when renewable energy is available while at the same time need to make sure that user QoS \color{black}requirements are honored. \color{black}
	
	\color{black}A fundamental problem of powering data centers with renewable energy sources is how to optimize the use of renewable energy. Powering data centers with renewable energy sources such as solar or wind is challenging as these sources of energy are non-dispatchable and are not always available due to their fluctuating nature. Thus, in this work, we aim to address this challenge through an optimization problem of \color{black}maximizing renewable energy usage while minimizing the usage of brown energy.\color{black}
	
	In this work, we address the research problem as: by predicting the amount of renewable energy, determining when to use brownout for interactive workloads and when to defer batch workloads, when to consolidate VMs to fewer hosts and scale hosts to maximize the usage of renewable energy while ensuring the QoS of workloads. Based on the detailed survey \cite{XuBrownoutSurvey}, this research problem has not been 
	addressed by previous work. 
	
	The key \textbf{contributions} of the paper are:
	\color{black}
	\begin{itemize}
		\item Provide a perspective model for multi-level adaptive resource scheduling to manage workloads and renewable energy;
		
		\item Propose a self-adaptive approach for interactive workloads and batch workloads to ensure their QoS by considering the predicted renewable energy at Denver city; 
		
		\item Implement a prototype system derived from the perspective model and the proposed approach on a small-scale  testbed;   
		
		\item Evaluate the performance of the self-adaptive approach in the proposed prototype system for web services. 
		
	\end{itemize}
	\color{black}

	The rest of the paper is organized as follows: Section \ref{sec:related} discusses the related work for managing energy in the cloud computing environment. Section \ref{sec: systemmodel} depicts the system model of our proposed approach,  
	followed by modeling and problem statement in Section \ref{sec:problemmodel}. 
	The scheduling algorithm with renewable energy is introduced in Section \ref{sec:algorithm}. Section \ref{sec:prototype} provides the detailed information about the implementation of our prototype system, and Section \ref{sec:performance} shows the evaluation results of our proposed approach under our prototype system. 
	Finally, conclusions along with the future directions are given in Section \ref{sec:conclusion}.

	\section{Related Work} \label{sec:related}
	\color{black}In this section, we discuss related research in the context of the dominant energy efficient approaches based on resource scheduling, brownout approaches, cooling-aware data center energy management and resource scheduling with renewable energy.   \color{black}
	\begin{table*}[]
		\caption{Comparison for related work}
		\label{tab:comparisonrelated}
		\resizebox{\textwidth}{!}{%
			\begin{tabular}{|c|c|c|c|c|c|c|c|c|c|c|c|}
				\hline
				\multirow{2}{*}{\textbf{Approach}} & \multicolumn{4}{c|}{\textbf{Technique}} & \multicolumn{3}{c|}{\textbf{Energy Model}} & \multicolumn{2}{c|}{\textbf{Workloads Type}} & \multicolumn{2}{c|}{\textbf{Resource Scheduling Layer}} \\ \cline{2-12} 
				& \textbf{DVFS} & \textbf{VM Consolidation} & \textbf{Host Scaling} & \textbf{Brownout} & \textbf{Brown Energy} & \textbf{Renewable Enegy} & \textbf{Cooling Energy} & \textbf{Single} & \textbf{Mixed} & \textbf{Single Layer} & \textbf{Multiple Layer} \\ \hline
				Beloglazov et al. \cite{Beloglazov} &  & $\surd$ & $\surd$ &  & $\surd$ &  &  & $\surd$ &  &  & $\surd$ \\ \hline
				Kim et al. \cite{Kim2011} & $\surd$ &  &  &  & $\surd$ &  &  & $\surd$ & \multicolumn{1}{l|}{} & $\surd$ &  \\ \hline
				Liu et al. \cite{liu2018thermal} & $\surd$ &  &  &  & $\surd$ &  & $\surd$ & $\surd$ & \multicolumn{1}{l|}{} & $\surd$ &  \\ \hline
				Teng et al. \cite{Teng2016} & $\surd$ & $\surd$ & $\surd$ &  & $\surd$ &  &  & $\surd$ & \multicolumn{1}{l|}{} &  & $\surd$ \\ \hline
				Nguyen et al. \cite{nguyen2017virtual} &  & $\surd$ & $\surd$ &  & $\surd$ &  &  & $\surd$ & \multicolumn{1}{l|}{} &  & $\surd$ \\ \hline
				Xu et al. \cite{xu2016energy} &  & $\surd$ & $\surd$ & $\surd$ & $\surd$ &  &  & $\surd$ & \multicolumn{1}{l|}{} & \multicolumn{1}{l|}{} & \multicolumn{1}{c|}{$\surd$} \\ \hline
				Hasan et al. \cite{hasan2017investigating} &  &  & $\surd$ & $\surd$ & $\surd$ & $\surd$ &  & $\surd$ &  &  & $\surd$ \\ \hline
				Li et al. \cite{li2018holistic} &  & $\surd$ &  &  & $\surd$ &  & $\surd$ & $\surd$ &  &  & $\surd$ \\ \hline
				Beloglazov et al. \cite{AntonTPDS} &  & $\surd$ & $\surd$ &  & $\surd$ &  &  & $\surd$ &  &  & $\surd$ \\ \hline
				Goiri et al. \cite{goiri2013parasol} &  &  & $\surd$ &  & $\surd$ & $\surd$ & $\surd$ &  & $\surd$ & $\surd$ &  \\ \hline
				Liu et al. \cite{liu2012renewable} &  &  & $\surd$ &  & $\surd$ & $\surd$ & $\surd$ &  & $\surd$ & $\surd$ &  \\ \hline
				Xu et al. \cite{Xu2019ICSOC} &  &  &  & $\surd$ & $\surd$ & $\surd$ & & $\surd$  & & $\surd$ &  \\ \hline
				Our Approach &  & $\surd$ & $\surd$ & $\surd$ & $\surd$ & $\surd$ & $\surd$ &  & $\surd$ &  & $\surd$ \\ \hline
			\end{tabular}%
		}
	\end{table*}
	
	\subsection{DVFS and VM consolidation}
	A large body of research on the energy efficiency of data centers has been dedicated to the optimization techniques to reduce the energy consumption of servers within a data center using technologies such as dynamic voltage and frequency scaling (DVFS) and VM consolidation~\cite{Beloglazov}\cite{Kim2011}. Liu et al~\cite{liu2018thermal} proposed a heuristic algorithm for big data task scheduling based on thermal-aware and DVFS-enabled techniques to minimize the total energy consumption of data centers. Kim et al. \cite{Kim2011} modeled real-time service as VM requests and proposed several DVFS algorithms to reduce energy consumption for the DVFS-enabled cluster.  
	Cheng et al. \cite{Cheng2018Energy} proposed a heterogeneity-aware task assignment approach to improve the overall energy consumption in a heterogeneous Hadoop cluster without sacrificing job performance. 
	Teng et al. \cite{Teng2016} presented a set of heuristic algorithms by taking advantage of DVFS and VM consolidation together for batch-oriented scenarios. 
	Nguyen et al. \cite{nguyen2017virtual} introduced a virtual machine consolidation algorithm with multiple usage prediction to improve the energy efficiency of cloud data centers. 
	
	Our work differs from these efforts in several perspectives: (1) none of these DVFS-based and VM consolidation approaches can function well if the whole system is overloaded; (2) none of them put the efforts to schedule the mixed type of workloads; (3) none of them applied the renewable energy to power their systems.
	
	\subsection{Brownout}

	Xu et al. \cite{XuBrownoutSurvey} proposed a survey and taxonomy on brownout-based approaches, which summarized the application of brownout in cloud computing systems for different optimization objectives.  Tomas et al. \cite{tomas2014straw} applied brownout to address the load balancing issues in clouds. 
	Shahrad et al. \cite{shahrad2017incentivizing} introduced a practical pricing model for brownout system and \color{black}aimed \color{black}to increase the utilization of the cloud infrastructure by incentivizing users to dampen their usage fluctuations. 
		\color{black} 
		Trade-offs exist when it comes to applying brownout and deactivating microservices. For example, the trade-off between energy and revenue is investigated in \cite{xu2017energy}. \color{black}In contrast, our optimization objective is managing the energy usage in cloud data centers. 
	
	Xu et al. \cite{xu2016energy}\cite{XuiBrownout2019} presented brownout-based approaches to manage microservices and resources from the energy perspective.    
	Hasan et al.~\cite{ hasan2017investigating} investigated the green energy and user experience trade-off in interactive cloud applications and proposed a controller to provide guarantees of keeping response time within the \color{black}Service Level Agreement (SLA) \color{black}range in the presence of green energy based on a brownout-enabled architecture. In contrast, this work advances the previous ones by: (1) managing the energy in a holistic way by considering multiple layers resource management, cooling power and mixed type of workloads; (2) incorporating renewable energy usage to reduce brown energy usage.
	
	
	\subsection{Holistic Management with Cooling}
	Due to the complexity of thermal modeling of data center operation, traditional approaches ignored the impacts of resource management techniques on the cooling power system of data centers. 
	Recently, the holistic management of resources in which both computing and cooling energy are considered in the minimization of the overall consumption of energy has gained considerable attention from the community. Li et al~\cite{li2018holistic}, for example,  provided models capturing thermal features of computer room air conditioning (CRAC) unit of the data center and accordingly propose a VM scheduling algorithm to reduce data center energy consumption while it maintains the SLA violation in an acceptable range. In their work, resource scheduling happens on VM level and the workload type is batch. 
	Al-Qawasmeh et al. \cite{al2015power} presented power and
	thermal-aware workload allocation in the heterogeneous cloud.  They developed optimization techniques to assign the performance states of CPU cores (P-states) at the data center level to optimize the power consumption while ensuring performance constraints. 
	Tang et al. \cite{tang2008energy} investigated the thermal-aware task scheduling for homogeneous HPC data center, which aims to minimize peak inlet temperature through task assignment, thus reducing the cooling power. 
	
	Unlike these efforts, our work considers the management of virtualized resources to optimize the resource usage. In addition, we consider the multiple layers resource scheduling and mixed types of workloads. 
	
	
	\subsection{Renewable Energy}
	Studies have been investigated in the literature that focused on the optimization of on-site renewable energy use in data centers.
	Goiri et al~\cite{goiri2013parasol} presented a prototype of a green data center powered with solar panels, a battery bank, and a grid-tie which they have built as a research platform. They also describe their method, called GreenSwitch, for dynamically scheduling the interactive and batch workload and selecting the source of energy to use. GreenSwitch aims to minimize the overall cost of electricity while respecting the characteristics of the workload and battery lifetime constraints. 
	In contrast, while their work focused on the resource scheduling at the application level, our work is a multiple layers scheduling approach that considers the application, VMs, and hosts. We also model the applications to be fitted into the brownout feature. In addition, our renewable energy prediction model based on support vector machine differs significantly from their  model \color{black}in which \color{black}solar energy was predicted based on the last epoch.   
	
	Liu et al.~\cite{liu2012renewable} also focused on shifting workloads and matching renewable energy supply and demand in the data center. They schedule non-critical IT workload and allocates IT resources within a data center according to the availability of renewable power supply and the efficiency of the cooling system. They \color{black}formulated \color{black}the problem as a constrained convex optimization and aim to minimize the overall cost within the data center. \color{black}Different from the optimization of the overall costs, we aim to optimize the energy perspective. Another difference is that we can optimize the power consumption by scheduling both interactive workloads and batch workloads, while \cite{liu2012renewable} only \color{black}optimizes \color{black}the scheduling of batch workloads without optimizing interactive workloads.  \color{black}
	
	
	To optimize onsite renewable energy use, in our previous work~\cite{Xu2019ICSOC}, we focused on microservices management problem as finite Markov Decision Processes (MDP). Similar to this work, the proposed method dynamically switches off non-mandatory microservices of the application to strike a balance between the workload execution and brown energy consumption. Following a greenness property value, it also suggests in what capacity the battery power should be consumed in each time slot. Different from that work, we consider joint management of both interactive workloads and batch workloads. We also cover the entire stack of resource scheduling including microservices, VMs and physical hosts. In addition, we test our system in a real testbed, while ~\cite{Xu2019ICSOC} only conducts simulation.

	The current paper contributes to the growing body of work in the related area. Table \ref{tab:comparisonrelated} summarizes the comparison among the related work based on the key techniques, energy model, workload types and resource scheduling layers. \color{black} Given the contributions of existing work, it is \color{black}important \color{black}to highlight the key difference between our proposed work and prior work. To be best of our knowledge, our work is the first to jointly manage interactive workloads based on brownout and batch workloads based on deferral approach. Prior work have only included one of these features. We also consider multiple layers resource scheduling based on microservices management, VM scheduling and host scaling with real testbed, which enables to form a truly integrated resource management. \color{black}

	\section{System model} \label{sec: systemmodel}
	
	\begin{figure}[!ht]
		\centering
		\includegraphics[width=1.0\linewidth]{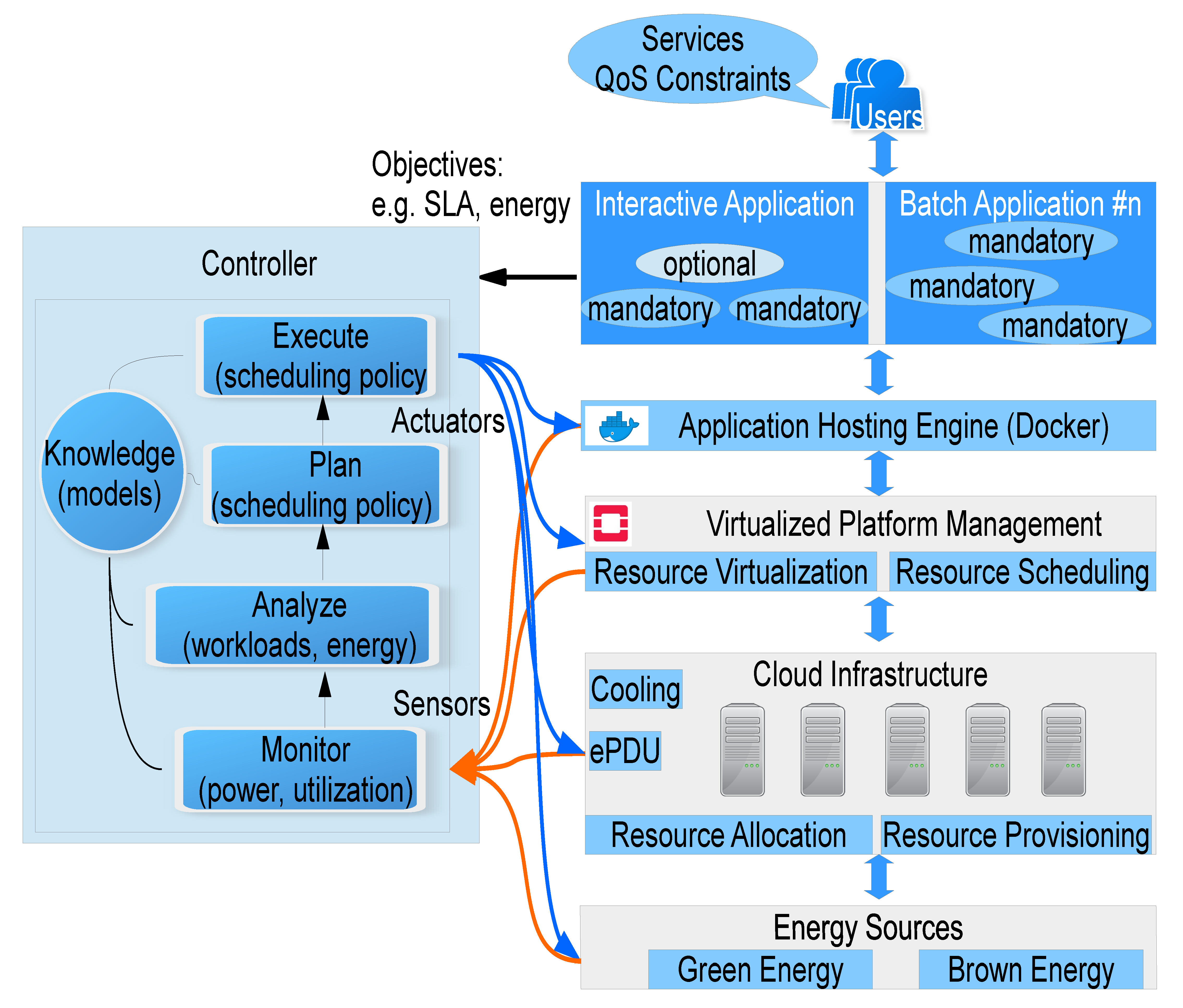}
		
		\caption[VarPerOptCom]{\color{black}Perspective Model}

		\label{images/perspectivemodel}
	\end{figure}
	
	
	\color{black}In this section, we propose a system model for adaptive resource scheduling as shown in Figure \ref{images/perspectivemodel}. We consider both interactive workloads and batch workloads in the application layer and consider green and brown energy together in the energy supply layer.  \color{black}
	
	In the users layers, users submit their service requests to the system. The users can define QoS constraints for the submitted requests, such as budget and deadline. The submitted requests are forwarded to the application layer. From the service providers' viewpoint, these workloads generated by users are processed by the applications hosted in the cloud infrastructure.   
	
	We consider two types of applications: the interactive application (such as web application), and batch application. The interactive application should be executed as soon as possible to ensure the QoS.  We consider the interactive application to support brownout, thus the microservices of the interactive applications can be identified as optional or mandatory. The optional ones can be deactivated to save resource usage, if deemed necessary. For the batch application, the workloads can be deferred for execution if their deadline is ensured. 
	
	Applications provided by service providers to offer services for users are managed by the application hosting engines, such as Docker \cite{DockerDoc} or Apache Tomcat. Applications can be deployed on either virtualized platform (virtual resources) or cloud infrastructure (physical resources). The host application engine can be container-based management platforms, e.g. Docker Swarm \cite{DockerSwarm}, Kubernetes \cite{Kubernetes} or Mesos \cite{Mesos}, which provide management of container-based applications. \color{black}The application containing multiple microservices can be deployed on multiple VMs. \color{black}The virtualized platform manages the virtualized resources, for example, the Virtual Machines managed by VMware \cite{VMware}. As for the resource allocation and provisioning in cloud infrastructure, they can be managed by infrastructure management platform, e.g. OpenStack \cite{OpenStack}. \color{black}Multiple VMs can be deployed on multiple hosts.\color{black}
	
	The bottom layer is the energy supply layer, which provides the mixed of energy sources for powering the system. The brown energy comes from a coal-based thermal power plant, which has  high carbon footprint. The green energy comes from the renewable energy, such as solar power.

	
	To support the resource provision, monitor and allocation in the system,   a controller is required based on MAPE-K \color{black}(Monitor, Analyze, Plan, Execute and Knowledge) \color{black} architecture model and fits into the feedback loop of the MAPE-K process, which has modules including \textit{Monitor}, \textit{Analyze}, \textit{Plan} and \textit{Execute} to achieve the adaptation process in cloud computing system \cite{Nallur2012}\cite{chen2017self}. Sensors and Actuators are used to establish interactions with the system. Sensors gather the information from different levels in the system, including application hosting engine, virtualized platform, cloud infrastructure, and energy usage. The sensors can be the devices attached to hardware, e.g. power meter. The collected information is provided to the Monitor module. 
	
	The Analyze module analyzes the received information from the Monitor module, and the Plan module makes decisions for applying scheduling policies, in which the scheduling policies are implemented. According to the decisions, the Execute module schedules resources via actuators on the application hosting engine and the virtualized platform to enable/disable optional microservices in interactive applications or defer the workloads of batch applications to be supplied by renewable energy. These operations can be fulfilled via the \color{black}Application Programming Interfaces (APIs) \color{black}provided by the application hosting engine or the virtualized platform. 
	
	The Knowledge pool in MAPE-K model is applied to store the predefined objectives (energy efficiency or SLA constraints) and trade-offs (e.g. trade-offs between energy and SLA). The rules in Knowledge pool, such as SLA rules, can be updated according to resource scheduling algorithms. The Knowledge pool also contains models like predicting the supplied amount of renewable energy, which can be used by scheduling algorithms. \color{black}. 
	
	In the following sections, we will introduce our proposed approach and the prototype system that is derived from this perspective model. 
	
	\section{Problem Modeling} \label{sec:problemmodel}
	\color{black}In this section, we will discuss our modeling and optimization problem, including the power consumption model, workloads model, and optimization objectives. \color{black}
	
	\subsection{Power Consumption}

	\subsubsection{Server Power Consumption}
	The server power model is derived from \cite{Zheng2014}, which is based on the average \color{black}CPU utilization\footnote{\color{black}In this work, we  consider the CPU utilization as the main source of resource consumption.}. \color{black}As we consider multiple layers scheduling, the utilization of hosts, VM and microservices are modeled:  
	\begin{equation}
	P_i^s = 	\begin{cases}
	P_i^{idle} + \theta_t\sum_{j=1}^{w_i} U_{i,j}^{vm} \times P_i^{dynamic}   & ,w_i > 0\\
	0 & ,w_i = 0
	\end{cases}
	\end{equation}
	where $P_i^s$ is the power consumption of the host $i$ in the data center, which is composed of two parts: the idle power $P_i^{idle}$ and dynamic power $P_i^{dynamic}$. The dynamic power part is related to the VM utilization on the host. If there is no VM running on the host, it means the host can be switched into the low power mode (S-State) and consume low energy. The $w_i$ represents the number of VMs deployed on host $i$. $U_{i,j}^{vm}$ represents the utilization of $jth$ VM on host $i$. The $\theta_t$ is the dimmer value of brownout at time interval $t$, which represents the percentage of resource utilization provisioned to the active microservices. For instance, if $\theta_t = 0.8$, it means 20\% utilization will be reduced by deactivating microservices. The dimmer value is calculated based on the number of overloaded hosts and more details will be given in Section \ref{sec:baiw}. 
	
	The utilization of VM is the sum of microservices utilization running on the VM, which is modeled as:
	\begin{equation}
	U_{i,j}^{vm} = \sum_{k=1}^{A_j} U_{j,k}^{ms}
	\end{equation}
	where the $ms$ is the id of microservice and $A_j$ is the number of microservices. Since CPU computation is main power consumption component of servers, in our server model, we mainly focus on the power draw by the CPU utilization.
	\subsubsection{Cooling Power Consumption}
	
	\color{black}
	We consider the data center thermal control is managed by Computer Room Air Condition (CRAC) system. The system contains multiple CRAC units, which transfer cold air to the hosts to reduce hotspots. Based on server power consumption and cooling efficiency, we can calculate the power consumed $P_i^c$ by cooling  equipment for host $i$ as:
	
	\begin{equation}
	P_i^c = \frac{P_i^s}{CoP(T_{sup})}
	\end{equation}
	There are some complex cooling models \cite{kaur2015energy}, while they are beyond the scope of this paper. We use the model from HP lab data center \cite{moore2005making} as follows:
	\begin{equation}
	CoP(T_{sup}) = 0.0068T_{sup}^2 + 0.0008T_{sup} + 0.458
	\end{equation}
	The $CoP$ is a function to estimate the cooling efficiency of cold air supply temperature  $T_{sup}$ provided by cooling equipment, which is related to the target temperature that room is aimed to be maintained. 
	\color{black}
	The total power draw by the server part and the cooling part can be represented as:
	\begin{equation}
	P_i = P_i^s+P_i^c
	\end{equation}
	The total power of data center with $k$ servers:
	\begin{equation}
	P_t = \sum_{i=1}^k P_i
	\end{equation}
	\subsection{Workloads Model}
	In this work, we consider two types of workloads: (1) interactive workloads and (2) batch workloads. The interactive workloads are response time sensitive, thus these workloads should be executed immediately with the response time specified in the SLA, while the batch workloads can be deferred for execution as long as the deadline is satisfied. 
	
	\color{black} Based on the different characteristics of these workloads, we assume that there are $M$ interactive workloads, and the amount of interactive workload $m$ at time $t$ is denoted as $a_m(t)$\color{black}, and consumes resource $u_m$\color{black}. This is a general model and the amount of workloads can be derived from analytical models (e.g. M/M/k) or realistic traces. \color{black}Thus, the demanded resources of interactive workload $m$ at time $t$ are $a_m(t) \times u_m$. Based on M/GI/1/PS model, the demanded resources can have a relationship with the target response time $rt_m$ as $\frac{1}{\mu_m-\lambda_m(t)/(a_m(t)\times u_m) } \leq rt_m$, where $\lambda_m(t)$ is the arrival rate and $\mu_m$ is the mean service rate. Thus, the minimum resources for interactive workloads to satisfy $rt_m$ is: $a_m(t)\times u_m = \frac{\lambda_m(t)}{\mu_m - 1/rt_m}$. \color{black}
	
	We also assume that there are $N$ types of batch workloads, type $n$ batch workloads have total demand $B_n$. Let $b_n(t)$ denote the amount of type $n$ batch workloads at time interval $t$, with start time $S_n$, execution time $E_n$, and deadline $D_n$, \color{black}consuming resource amount $u_n$. \color{black}We use $b_n(t)^{'}$ to denote the original amount of batch workloads, the actual amount of workloads should add the workloads deferred from the previous time slot and minus the workloads that will be deferred to the next time slot. We use $\gamma_n^{t-1}$and$ \gamma_n^{t}$ to present the percentage of deferred workloads at time interval $t$ and $t-1$ for type $n$ batch workloads. Then we have 
	$b_n(t) =  \gamma_n^{t-1} \times b_n(t-1) +b_n(t)^{'} - \gamma_n^{t} \times b_n(t)^{'}$.  
	
	Therefore, the total \color{black}CPU resource demands \color{black}at $t$ are: \color{black}
	\begin{equation}
	d(t) = \sum_{m} a_m(t)\times u_m+\sum_{n}b_n(t) \times u_n
	\end{equation}
	The value of $d(t)$ should be $0 \leq d(t) \leq D$, in which $D$ is the maximum \color{black}CPU \color{black}resource capability of the system. \color{black} \color{black} To be noted, rather than constant, $d(t)$ is varied based on our proposed scheduling policy, for instance, \color{black}CPU resource \color{black}provisioned to interactive workloads can be adjusted by brownout mechanism (Equation (1)) and the batch workloads can be deferred based on system status (Equation (7)). 
	
	
	\color{black}
	

	\subsection{Optimization Objectives}
	\color{black}
	We assume the scheduling period as $T$, and the time interval at which we schedule resources is denoted as $t$. We assume the available renewable energy at time $t$ is $R_t$, which can be predicted by prediction approaches, e.g. machine algorithms. \color{black}As the server power and cooling power is related to workloads, we use \color{black}$d(t)^{'}$ \color{black}to denote the power consumption resulted from the workload execution on servers, and \color{black}$c(d(t)^{'})$ \color{black}represents the cooling power resulted from the workloads, thus \color{black}$P_t = d(t)^{'} + c(d(t)^{'})$. \color{black}Our optimization objective is modeled as: 
	\begin{equation}
	\begin{aligned}
	min \sum_t (max(P_t - R_t, 0))  \\
	s.t. \quad 0 \leq d(t) \leq D, \ \forall t\\
	0 \leq \theta_t \leq 1, \ \forall t\\
	b_n(t) =  \gamma_n^{t-1}  \times b_n(t-1) +b_n(t)^{'} - \gamma_n^{t} \times b_n(t)^{'}, \ \forall t \\
	\sum_t b_n(t) \leq B_n, \ \forall n \\
	\end{aligned}
	\end{equation}
	where aims to minimize the usage of brown energy by maximizing the usage of renewable energy. \color{black} Our proposed solution makes schedule decisions of each time interval slot for interactive workloads and batch workloads based on the availability of renewable energy. \color{black} Meanwhile, the constraints including maximum capacity $D$ in the system, dimmer value $\theta_t$,  maximum total demand of batch workloads should be satisfied. \color{black}As this optimization objective is convex in $d(t)$, so it can be solved efficiently. \color{black}
	

	\section{Scheduling with renewable energy} \label{sec:algorithm}
	\color{black}In this section, based on the problem modelling, we introduce our proposed scheduling algorithm with renewable energy for both interactive workloads and batch workloads.   \color{black}

	\subsection{Green-aware Scheduling Algorithm} \label{sec:gasa}
	\begin{algorithm}[t]
		\color{black}
		\footnotesize
		\caption{Green-aware scheduling algorithm \label{gasa}}
		\SetAlgoLined
		\SetKwInOut{Input}{Input}\SetKwInOut{Output}{Output}
		\SetKwFor{ForAll}{forall }{do}{end forall}	
		
		\Input{host utilization $U_i^t$, utilization thresholds $TU_{up}, TU_{low}$}
		\Output{brown energy usage $\sum_t(max(P_t - R_t, 0))$} 
		$n_o^t = \sum_{k}(U_i^t > TU_{up})$ \\
		
		\For{$t \leftarrow 0$ to $T$}
		{
			
			\uIf{$n_o^t > 0$}
			{
			   \color{black}	\tcp{take actions to minimize $\sum_t(max(P_t - R_t, 0))$ } \color{black}
				$S_i$ $\leftarrow$ brownout algorithm for interactive workloads \\
				$T_j^{D} \leftarrow$ deferring algorithm for batch workloads \\
			}
			\uElseIf{$U_{avg}^t < TU_{low}$}
			{   \color{black}\tcp{take actions to minimize $n_a$}  \color{black}
				run VM consolidation algorithm \\ 
				run host scaling algorithm \\
		}
			\Else{
				process iteractive workloads in normal mode\\ 
				porcess batch workloads in normal mode\\
			}
		}
		
	\end{algorithm}
	
	To schedule the interactive and batch workloads in an energy efficient manner by considering renewable energy, we propose a Green-aware scheduling algorithm, which is shown in Algorithm 1. During the observation period $T$, at each time interval $t$, the algorithm will firstly identify the number of overloaded hosts (line 1). If the overloading situation exists, the algorithm will manage the interactive workloads and batch workloads with different algorithms: brownout algorithm for interactive workloads (Algorithm 2) and deferring algorithm for batch workloads (Algorithm 3) to minimize brown energy usage (lines 4-5). \color{black}These two algorithms will do the actions including deactivating microservices and deferring workloads to achieve the brown energy usage objective. 
	Here we assume the interactive workloads \color{black}utilize more CPU utilization \color{black}than the batch workloads, so the interactive workloads are processed earlier to achieve better scheduling effects. 
	If the system is not overloaded and the average utilization is below the underutilized threshold (line 6), the algorithm will apply VM consolidation algorithm derived from \cite{Beloglazov} (line 7) that consolidates VMs to the hosts that produce the minimum incrementation of energy consumption, and apply host scaling algorithm (Algorithm 4) to change the number of active hosts $n_a$. The motivation is that the idle servers will be switched into the low power mode to save energy (lines 7-8). If the system is running at the normal status, then the workloads will be executed in the normal fashion.
	\color{black}    
	
	\subsection{Brownout Algorithm for Interactive Workloads} \label{sec:baiw}
	
	\begin{algorithm}[t]
		\color{black}
		\footnotesize
		\caption{Brownout Algorithm for Interactive Workloads \label{BrownoutIW}}
		\SetAlgoLined
		\SetKwInOut{Input}{Input}\SetKwInOut{Output}{Output}
		\SetKwFor{ForAll}{forall }{do}{end forall}	
		
		\Input{time interval $t$, the number of overloaded hosts $n_o^t$, the percentage of utilization from batch workloads $\epsilon$, \color{black}and the starting time and end time of the available renewable energy $t_r^s$, $t_r^e$}
		\Output{deactivated microservices $S_i$} 
		
		\For{host $i$ in the host list}{
			\eIf{$ t < t_r^s$ $||$ $t > t_r^e$}
			{
				$\theta_t = \sqrt{\frac{n_o^t}{n}}$ \\
				$U_i^r = \theta_t \times U_i^t$ \\
				\If{$U_i^t > TU_{up}$  }
				{	
					\color{black}\tcp{take actions to minimize $|U_i^r - U(S_i)|$} \color{black}
					find deactivated microservices $S_i$ on host $i$ \\
					deactivate the microservices}
			}
			{
				\If{$ R_t<P_t$}
				{
					$\theta_t = \frac{1}{1-\epsilon} \times \sqrt{\frac{R_t}{P_t}}$ \\
					$U_i^r = \theta_t \times U_i^t$ \\
					\color{black}\tcp{take actions to minimize $|U_i^r - U(S_i)|$} \color{black}
					find deactivated microservices $S_i$ on host $i$ \\
					deactivate the microservices \\
				}
			}
		}	
		
	\end{algorithm}
	The pseudocode of the brownout algorithm for interactive workloads is shown in Algorithm 2. The algorithm schedules resources differently according to whether the renewable is available or not. \color{black}The starting time and end time of available renewable energy are denoted as $t_r^s$ and $t_r^e$ respectively. \color{black}1) During the time when renewable energy is not available (line 2), the brownout is triggered, and the dimmer value is generated. The dimmer value $\theta_t$ is computed based on the severity of overloads in the system (line 3). With the dimmer value, the expected utilization reduction $U_i^r$ on host $i$ is computed (line 4). Then the algorithm selects a set of microservices $S_i$ to deactivate, thus the utilization is reduced. The difference between the expected utilization reduction $U_i^r$ and the sum of utilization of selected microservices $U(S_i)$ is minimized (lines 6-7). \color{black}In lines 6-7, to minimize the difference, the microservices selection process sorts the microservices according to their utilization in a list,  and finds the sublist which has the utilization that is closest to $U_i^r$. \color{black}To be noted, the selection process will only search for the optional microservices, which means if there are not enough or no optional microservices for deactivation, only the available or no microservices will be selected. \color{black} 2) When the renewable energy is available but less than the total required energy, the brownout is also triggered (line 10). The dimmer value is calculated based on the renewable energy $R_t$ and required energy $P_t$ as noted in line 11. \color{black}Then the remaining steps are the same as in the first part of Algorithm 2, which finds the microservices and deactivates them (same as in lines 6-7). \color{black}3) When sufficient renewable energy is available, brownout will not be triggered.

	\subsection{Deferring Algorithm for Batch Workloads} \label{dabw}
	\begin{algorithm}[t]
		\color{black}
		\footnotesize
		\caption{Deferring Algorithm for Batch Workloads \label{AlgIB}}
		\SetAlgoLined
		\SetKwInOut{Input}{Input}\SetKwInOut{Output}{Output}
		\SetKwFor{ForAll}{forall }{do}{end forall}	
		
		\Input{batch workload $b_n(t)$ with start time $T_n$, execution time $E_n$, deadline $D_n$, \color{black} and the starting time and end time of the available renewable energy $t_r^s$, $t_r^e$}
		\Output{deferred time $T_n^d$} 
		\For{$t=T_n$ in $b_n(t)$}{
			\uIf{\color{black}$0 <t<t_r^s$}{
				\eIf{$D_n < t_r^s$}
				{execute $b_n(t)$ }
				{defer $T_n^d$ time for execution \\
					$td = t+ T_n^d$, $\forall$ $td \leq D_n -E_n, td > t_r^s$ \\
					$d(td)^{'} = \sum_{m} a_m(td)^{'}+\sum_{n}b_n(td)^{'}$ \\
					$R_{td}^{'} > P_{td}^{'}$ \\
					$T_n^{'} = td$ \\ 
					update $P_{td}^{'}$}
				
			}
			\uElseIf{\color{black} $t_r^s \leq  t \leq$ $t_r^e$}{
				\eIf{$R_t > P_t$}
				{execute $b_n(t)$  }
				{defer $T_n^d$ time for execution \\
					$td = t+ T_n^d$, $\forall$ $td \leq D_n -E_n$ \\
					$d(td)^{'} = \sum_{m} a_m(td)^{'}+\sum_{n}b_n(td)^{'}$ \\
					$R_{td}^{'} > P_{td}^{'}$ \\
					$T_n^{'} = td$ \\ 
					update $P_{td}^{'}$}
			}
			\Else{
				execute $b_n(t)$ 
			}
		}

	\end{algorithm}
	
	Algorithm 3 shows pseudocode for processing the batch workloads. The batch workloads are executed when their start time $S_j$ is coming (line 1). The workloads are processed based on the time period that the workloads are in. For the workloads which have the start time before the renewable energy start time $t_r^s$, the objective is to defer their execution to the time when the renewable energy is available while ensuring their deadlines (lines 2-12).  1). If the deadline is before $t_r^s$, it means the workload cannot be deferred to be processed by renewable energy, so the workload can be executed at $t$ (lines 3-4). If the workload can be deferred, the algorithm defers its time with $T_n^d$, then the algorithm updates the workloads at time $td$, which equals to $t+T_n^d$. \color{black}The deferred time $T_n^d$ should satisfy the constraint, e.g. not failing the deadline, and the renewable energy is enough at $td$. \color{black}If the constraints are satisfied, the algorithm updates the predicted power consumption at $td$. 2). When the start time of the workload is during the time when renewable energy is available and sufficient, the workload is executed; otherwise, the workload will be deferred \color{black}(lines 13-23)\color{black}. Similar to the first part of Algorithm 3, the deferred time $td$ also needs to satisfy the constraints in Equation (8). 3) When the time is after $t_r^e$, it means the renewable energy is not available any more, therefore, the workloads are executed as soon as possible to comply with the deadlines \color{black}(line 25)\color{black}. 
	
	\subsection{Host Scaling} \label{hs}
	
	\begin{algorithm}
		\color{black}
		\footnotesize
		\caption{Hosts scaling algorithm \label{hostscale}}
		\SetAlgoLined
		\SetKwInOut{Input}{Input}\SetKwInOut{Output}{Output}
		\SetKwFor{ForAll}{forall }{do}{end forall}	
		\Input{number of hosts  $n$ in data center, 
			number of requests when  host is overloaded $num_{thr}$, predicted number of requests $\hat{num}(t)$ at time $t$.} 
		\Output{number of active hosts $n_{a}$ }
		$n_{a} \leftarrow \lceil \hat{num}(t) \div num_{thr} \rceil $    \\
		$n'  \leftarrow n_a - n $

		\uIf{$n' > 0$}{
			Add $n'$ hosts \\
			\While{$P_t \leq R_t$}
			{Add another host \\
				update $P_t$}	
		}
		\uElseIf{$n' < 0$}{
			Remove $|n'|$  hosts  $\quad$$\quad$//$|n'|$ is the absolute value of $n'$
		}
		\Else{
			no host scaling
		}
		\textbf{return} $n_{a}$
		
	\end{algorithm}
	We use a modified host scaling algorithm from \cite{Toosi2017} by considering renewable energy as shown in Algorithm 4. With profiling data, we configure the threshold of requests that leads to overloads, in which the average response time violates the predefined constraints. \color{black}The predicted number is calculated based on the number of requests in recent time slots derived from the prediction approach in \cite{XuiBrownout2019}. \color{black}
	The algorithm calculates the difference $n^{'}$ between the number of required servers and actual servers.  1). When more servers are needed, then it adds $n^{'}$ servers into the system (lines 3-4). If the renewable energy is still enough, then it tries to scale more servers into the system to improve the QoS (lines 5-8). 2). If servers are already enough, then remove $|n^{'}|$ servers from system to reduce energy. 3). If $n^{'}$ is 0, then it means no host scaling is required. 
	
	\subsection{Renewable Energy Prediction} \label{sec:renewprediction}
	\color{black}
	
	\color{black}In practice, green data centers with onsite solar installations or wind farms are powered by electricity generated from the renewable source while they are backed up with the Grid (they use inverters that automatically switch energy source based on the availability). In some other potential scenarios, they are just connected to the Grid and pay for electricity generated from renewable energy sources. In both scenarios, physical servers (electricity consumers) are oblivious to energy sources. Therefore, in this work, we assumed that servers use electricity generated from renewable energy sources without compromising reality and precision. As per our assumption, if renewable energy is insufficient or not available, Grid (Brown) electricity will be used. \color{black}
	
	We focus on the solar energy as it is one of the most common sources of renewable energy. 
	We use Support Vector Machine (SVM) to predict the solar irradiation or PV power output for the availability of renewable energy, which is a machine learning approach \color{black}that has been applied to data analysis successfully. In the studies related to solar irradiation prediction  \cite{bae2017hourly}\cite{belaid2016prediction}, SVM has been used to forecast and train the solar radiance model. \color{black}
	
	Since we do not have the access to the hourly solar irradiance at Melbourne City, in this paper, we use the historical data from NREL Solar Radiation Research Laboratory \footnote{ https://midcdmz.nrel.gov/srrl\_bms/}. The solar panels of the Laboratory are located at Denver, Colorado, US (Latitude $39.742^o$ North,  Longitude $105.18^o$ West), which has a similar weather to Melbourne instead. 
	We use the hourly-based solar irradiance data from September 1 2018 to November 1 2018. 
	SVM prediction approach has two phases: the training phase and testing phase. 80\% data is used for the training phase, and 20\% data is used for the testing phase. Once the process is finished, the test data and prediction results are compared to calculate the error rate. We use the SVM R toolbox for our purpose.

\begin{figure}[!ht]
		\centering
		\includegraphics[width=1.0\linewidth]{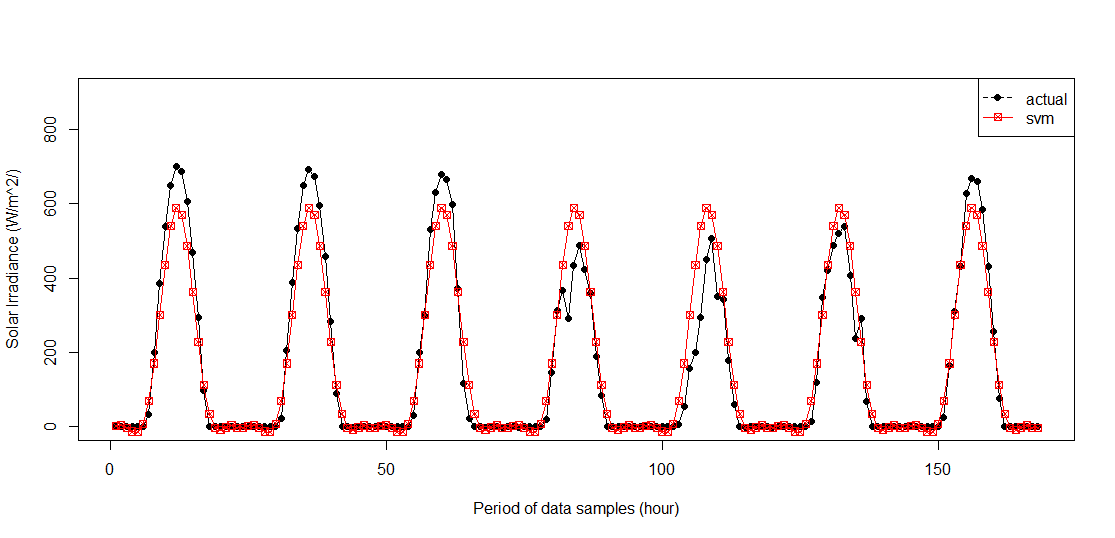}
		
		\caption[VarPerOptCom]{Denver Solar Radiation}

		\label{images/svm}
	\end{figure}

	The obtained results are shown in Figure \ref{images/svm}. It shows that the SVM-based approach can achieve the values close to actual ones. In the testing phase, the coefficient of determination ($R^2$) is 0.763 and correlation coefficient ($r$) is 0.873. The selected parameters for SVM are regularization parameter $C=4$ and Kernel bandwidth $\epsilon=0$. In the experiments section, we applied this trained model to predict solar irradiance. The solar irradiance can be easily calculated with the conversion efficiency of solar panels, e.g. \color{black}20\%. We \color{black}assume all solar irradiance will be available as power energy. \color{black} We use this trained model to predict the available renewable energy in advance for the observation period $T$, e.g. one day. \color{black}

	\section{Prototype System Implementation} \label{sec:prototype}
\begin{figure}[!ht]
		\centering
		\includegraphics[width=0.99\linewidth]{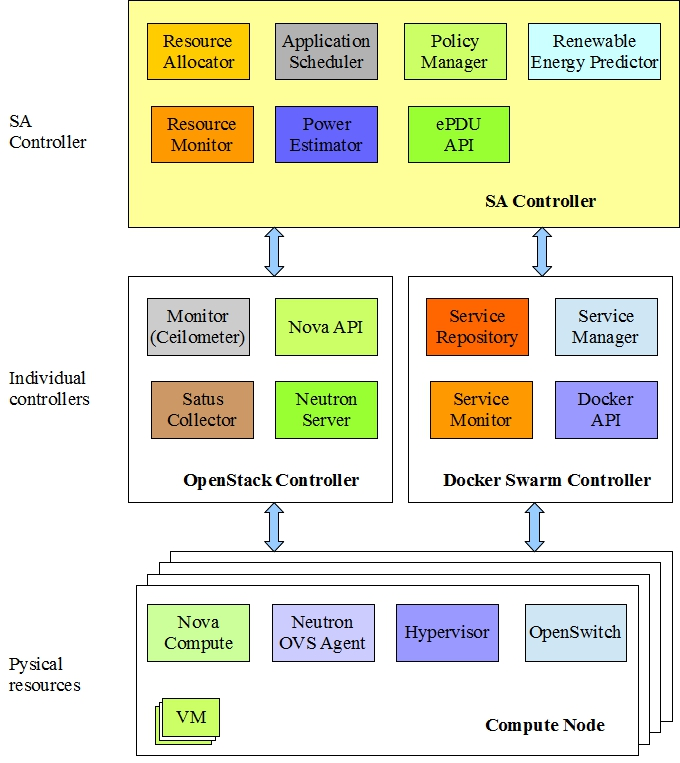}
		\caption[VarPerOptCom]{System architecture and underlying software components of prototype system}
		
		\label{images/sysarchi}
	\end{figure}

	To realize our system model in Section \ref{sec: systemmodel} and evaluate our proposed approach, we configure our testbed to develop a prototype system. Figure \ref{images/sysarchi} shows the implemented architecture of our prototype system. Cloud resource management platform and microservices management platform have been developed and widely used for years, thus, in this work, we design and implement our prototype based on these mature platforms. 
	
	\color{black}Cloud IaaS resource management platform, OpenStack, is responsible for managing cloud resources, including CPU, memory, and bandwidth. The monitored data of resources is collected by status collector and can be used for resource provisioning and optimization. The microservice management platform, Docker Swarm, is responsible for managing service images, monitoring service resource utilization and managing the service life cycles. Other Docker APIs can also be used to run operations on services. These two platforms are mapped to the Virtualized Platform Management layer and Application Hosting Engine layer respectively in the system model in Figure \ref{images/perspectivemodel}.
	
	Based on the two management platforms for cloud resources and services, SA (Self-Adaptive) controller is designed to manage and monitor both of them to achieve the multiple level resource scheduling, which is mapped to the Controller component in Figure \ref{images/perspectivemodel}. When requests are submitted to the system, like interactive workloads or batch workloads, the resource allocator in SA controller manages cloud resource management platform and service management platform simultaneously to accept and process requests by providing the requested amount of resources. Apart from allocating resources to requests, the resource allocator can also optimize resource utilization. For instance, brownout can be triggered to deactivate optional microservices to reduce resource utilization. The service provider can also configure the available resource scheduling policies for the energy efficiency purpose. 
	\color{black}

	To provision and optimize the resources by means of resource allocator, the resource monitor needs to collect the resource usage at different levels, including services utilization, VMs utilization, and hosts utilization. To minimize the overheads of frequently monitored data collection, the collection time intervals should be well configured by the service provider. For instance, the brownout mechanism can be checked every five minutes as the brownout costs are not high, while the VM migration and host scaling operations can be executed with longer time intervals, e.g. one hour. 
	
	In the following subsections, we introduce the implementation of our prototype system in details. 
	
	\subsection{Implementation}
	
	To implement our prototype system, we take advantage of the OpenStack cloud resource management platform and Docker Swarm service management platform. The system is implemented with Java, OpenStack, Docker Swarm, Ansible, Eaton Power Distribution Units (ePDU) API. Our prototype system uses these open source tools to provide a self-adaptive approach to optimize, manage and provision resources for different types of workloads. 
	
	OpenStack platform is used to manage hosts and VMs. The hosts in OpenStack are called compute nodes and are running with Nova Compute Node component to connect the hypervisor and OpenStack controller. VMs are managed by Nova API to create, migrate and remove VM instances. The Neutron OVS Agent and OpenVSwitch are providing services related to the network. 
	
	Docker Swarm Platform manages the service provided by service providers. The images of services are stored in the service repository component, which can fetch the images from remote to local. The services are managed by the service manager via Docker APIs, including creation and deletion. The status of services are monitored by a service which monitors service utilization and liveness.
	
	Our prototype system is based on these services to manage the resources and services to handle the requests. Below, we introduce the details of the components in our prototype.

	
	\textbf{Resource Allocator:} 
	It interacts with OpenStack controller via OpenStack APIs and Docker Swarm Controller via Docker APIs. It manages the physical resources on compute nodes, and these physical resources can be used for creating and deploying VMs on the nodes. Resource Manager knows the amount of resource that is used or remaining on each compute node, like the number of cores, memory, and storage. When creating a VM instance, it can also specify the instance capacity (CPU, memory, operation system and etc.) as well as other information related to VMs, such as location, images of VMs and IP address. The virtual network in a compute node is also managed by Resource Manager that uses the Neutron component, which is deployed on each compute node. 
	
	\textbf{Resource Monitor:} It is used to monitor the running status of the whole system from different levels, including hosts, VMs and services. We use OpenStack Ceilometer and Gnocchi components to measure the data at the host and VM level. Ceilometer is responsible for monitoring the utilization of resources of VMs and then sends the collected data to Gnocchi to aggregate the data for all the hosts. We use Docker APIs to collect the resource utilization of services deployed on VMs. Apart from monitoring the resource utilization, we also use ePDU APIs to monitor the power consumption of hosts. With these monitored data, other components, like Power Estimator and Policy Manager can use these data to make decisions, which will be introduced later. 
	
	\textbf{Application Scheduler:} We design our main controls in the Application Scheduler component. When requests are submitted by users, the Application Scheduler decides which requests in the batch workloads should be deferred, which microservice should be temporarily deactivated by brownout mechanism,  which VM should be migrated to another host and which host should be switched to the low power mode. With the retrieved data from the Resource Monitor component, these decisions are made with the policies in the Policy Manager. After the decisions are made, Resource Provisioner exploits Resource Manager to allocate the resources to VMs, services, and requests.

	\textbf{Power Consumption Estimator:} 
	To achieve our objective of managing energy and support our scheduling policies, we have a power consumption estimator to predict the power consumption at a specific time period. For example, for the batch workloads, we proposed a deferring algorithm, thus we need to estimate the power consumption at the deferred time period to calibrate our algorithm. We use the workloads model shown in Equation (7) to estimate the workloads and then convert it to the total energy consumption based on the model in \cite{liu2012renewable}. 
	
	\textbf{Policy Manager:} It contains the implemented scheduling policies in our prototype, e.g. Algorithms \ref{gasa} to \ref{hostscale}. The Policy Manager component uses the retrieved data from Resource Monitor, and makes decisions based on system status. For example, a VM is migrated from an underutilized host to other hosts, thus the idle host can be switched to the low power mode to save power consumption; when the renewable energy is not sufficient and the system is overloaded, to ensure the QoS of service, brownout can be triggered to relieve the overloaded situation. The customized workloads processing policy, VM migration policy and host scaling policy can also be implemented for the policy manager. 
	
	\textbf{ePDU API:} \color{black} 
	Eaton Power Distribution Units (ePDU) \footnote{https://powerquality.eaton.com/ePDUG3/} is an effective power management and monitoring device. It has outlets that \color{black}allow electric devices to \color{black}be connected to it. It also provides the features to read the power consumption of hosts as well as turn on/off the outlets remotely. We implemented Python scripts based on ePDU APIs to read the power data at per second rate to support part of the functions in Resource Monitor. Our scripts can also operate the hosts remotely by turning on/off the power supply to hosts to support the decision in Policy Manager. For example, a host needs to be scaled out if the whole system is underutilized; or hosts should be scaled in to support more requests. 
	
	\textbf{Renewable Energy Predictor:} 
	For supporting our renewable energy experiments, we implement a renewable energy predictor that predicts the renewable energy at Denver city based on the historical data. 
	As introduced in Section \ref{sec:renewprediction}, our prediction models show that it can achieve a high accuracy. The data based on this component can also be incorporated into the scheduling policy. 

	\section{Performance Evaluation} \label{sec:performance}
	\color{black}To evaluate the performance of our proposed approach, we conduct experiments in our implemented prototype system. We first present the environment settings in Section 7.1, and then introduce the workload and application settings in Section 7.2 and 7.3. The results are demonstrated and discussed in Section 7.4.   \color{black}
	\subsection{Environmental Settings}
	\begin{table}[]
		\caption{Machines Specification}
		\label{MachinesSpecification}
		\resizebox{0.5\textwidth}{!}{%
			\begin{tabular}{|c|c|c|c|c|c|c|}
				\hline
				\textbf{Machine}                          & \textbf{CPU} & \textbf{Cores} & \textbf{Memory} & \textbf{Storage} & \textbf{Idle Power} & \textbf{Full Power} \\ \hline
				3 $\times$ IBM X3500 M4                   & 2 GHz        & 12             & 64 GB           & 2.9 TB           & 153 Watts           & 230 Watts           \\ \hline
				4 $\times$ IBM X3200 M3                   & 2.8 GHz      & 4              & 16 GB           & 199 GB           & 60 Watts            & 150 Watts           \\ \hline
				2 $\times$ Dell OptiPlex 990 & 3.4 GHz      & 4              & 8 GB            & 399 GB           & 26 Watts            & 106 Watts           \\ \hline
			\end{tabular}%
		}
	\end{table}
	
	%
	%
	%
	\color{black}In our experiments, the upper utilization threshold $TU_{up}$ and lower utilization threshold $TU_{low}$ are configured as 80\% and 20\% respectively, as these values have been evaluated in our previous work \cite{XuiBrownout2019}\cite{Beloglazov} that they can achieve good trade-offs between energy consumption and QoS than other values. \color{black}For example, configuring the upper utilization threshold to be lower than 80\% can trigger  brownout too frequently, while the upper utilization threshold with 90\% or higher can hardly trigger brownout. \color{black}We also configure the scheduling time interval as 5 minutes and the whole scheduling period as one day.\color{black}

	\textbf{Hardware:} We utilize a micro data center of Melbourne CLOUDS lab as testbed. Our data center \color{black}consists \color{black}of 9 heterogeneous servers.
	Table \ref{MachinesSpecification} shows the capacity specification of the servers and their energy consumption information.  To monitor the power consumption of individual machines, we use two ePDUs and all the servers are connected to them. Apart from the power monitor, the ePDUs also enable us to switch on/off power outlets connected with individual server remotely through the network. 
	The total maximum power of the IT equipment in our system is 1.27 kWh for 8 hosts (one IBM X3500 M4 machine is regarded as the OpenStack control node and is not considered). 
	
	We assume our system is equipped with 1.63 \color{black}kW \color{black}PV panel\footnote{The total power of PV panels can be increased by adding more panels.}, \color{black}which has 30\% more power than the maximum power of hosts, \color{black}as the cooling part normally consumes about 20\% to 30\% of server energy if the target temperature is 25 degree \cite{Liu2012}. This cooling power consumption percentage is validated in the prototype system in \cite{goiri2013parasol}. \color{black}We consider to control the data center temperature as 25 degree, according to Equation (4), $T_{sup}=25$, then we get $CoP(T_{sup})=4.728$. In the following experiments, we use this value to compute the power from the cooling equipment, e.g. if the hosts consume 10 kWh, then the cooling part is 2.11 kWh. 
	
	
	\textbf{Software:}  The operating systems of the servers are CentOS Linux Distribution. We use OpenStack \cite{OpenStack} to support our cloud platform and manage the VMs. One of our most powerful machines is selected as our controller node, and other nodes are acting in the same role. In VM instances, we deploy Docker \cite{Docker2017a} containers to provide services in the form of microservices and use Docker Swarm to manage the containers cluster. Some other required software, like Java, Ansible are also installed in the VMs.

	\subsection{Workloads}
	To make the experiments as realistic as possible, we use real traces from Wikipedia and Facebook. For the interactive workloads, we use the real trace from Wikipedia requests
	on 2007 October 17 to replay the workload of Wikipedia users. The trace includes data
	on requests time stamp and their accessed pages. We filter the requests based on per
	second rate and generate the requests rate. The original request rate is around 1,500-
	3,000 per second. We use 10\% of the original requests and these requests can consume up to 43\% cluster utilization.
	
	For the batch workloads, we use the traces collected in October 2009 by Facebook for applications that are executed under Hadoop environment\footnote{https://github.com/SWIMProjectUCB/SWIM/wiki/Workloads-repository}. Referring to \cite{goiri2013parasol}, we configure the map phase of each job takes 25-13000 seconds, and the reduce phase takes 15-2600 seconds. The deadline for processing jobs is generated based on uniform distribution with $\mu=6$ hours and $\sigma=1$ hour in $N(\mu, \sigma^2)$. We also assume the workloads consume the maximum of cluster utilization as 27\% as same as in \cite{goiri2013parasol}. 
	
\begin{figure}[!ht]
		\centering
		\includegraphics[width=0.99\linewidth]{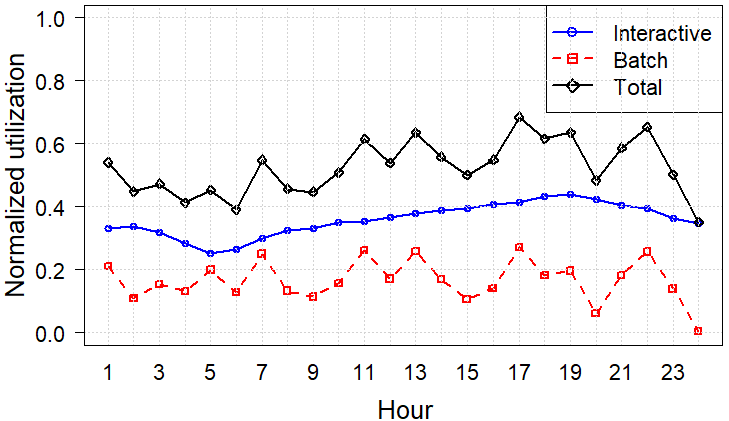}
		
		\caption[VarPerOptCom]{Workloads distribution}

		\label{fig:workloaddistribution}
	\end{figure}

	Figure \ref{fig:workloaddistribution} shows the one-day normalized resource utilization trace of the aforementioned workloads. We can clearly see the variance of utilization demand of both interactive and batch workloads, thus the workloads can be managed to fit into the availability of green energy. For instance, at hour 7, if the green energy is not sufficient to supply the all the workloads, then some batch workloads can be deferred to a later time when more green energy is available. 
	
	\subsection{Application}
	\color{black}
	We use the Weave Shop\footnote{See https://github.com/microservices-demo/microservices-demo for more details.} web application that is implemented with containers as the application to process the interactive workloads derived from Wikipedia traces. The Weave Shop is a web-based shopping system for selling socks online and has multiple microservices, including user microservice to handle user login, user database microservice for user information storage, payment microservice to process transactions, front-end microservice to show the user interface, catalog microservice for managing item for sale etc. As these microservices are implemented independently, they can be deployed and controlled without impacting other microservices. The application is deployed by a configuration file, and 30\% of the microservices are configured to be optional, e.g. the recommendation engine. \color{black}The generated workload is communicated to the mandatory microservices (e.g. the font-end microservices) to measure the response time. \color{black}The microservices are deployed on the cluster with multiple VMs. The head node of the cluster is deployed with a gateway microservice in Weave Shop that is responsible for distributing the interactive workloads to different microservices deployed on multiple VMs. The application and VMs are deployed on the hosts with dynamically adjusted. \color{black}
	\subsection{Results}
	To evaluate the benefits of our proposed approach for renewable energy usage, we perform the comparison of our proposed Green-aware and Self-adaptive approach (GSA) and a state-of-the-art baseline algorithm (HS), \color{black}which applies VM consolidation based on Modified Best Fit Decreasing  algorithm \cite{Beloglazov} that consolidates VMs to the hosts that produce the minimum energy incrementation, and host scaling \cite{Toosi2017} that dynamically adds/removes hosts in system based on profiling and workloads prediction. \color{black}
	
	\begin{figure*}[t]
		\centering
		\begin{subfigure}{0.48\linewidth}
			\centering
			\includegraphics[width=0.99\linewidth]{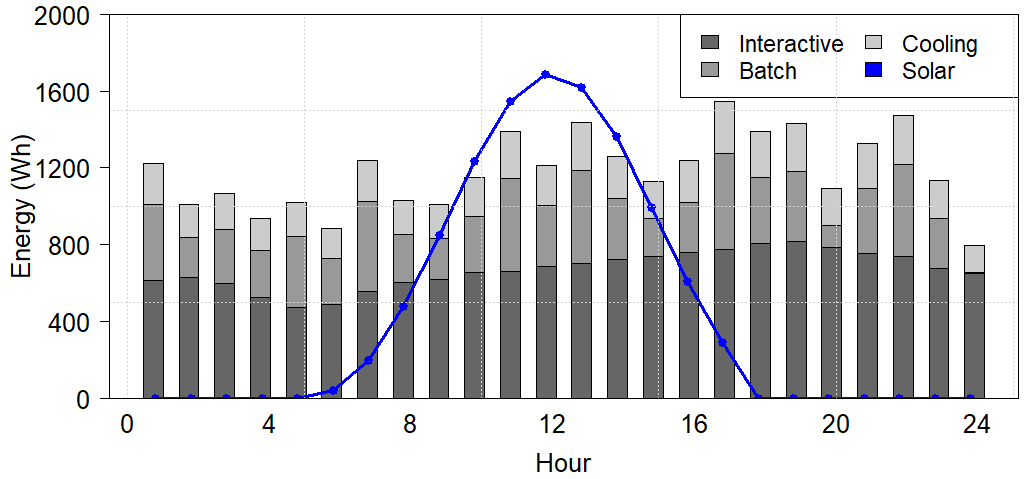}
			\caption{}
			\label{images/noSA}
		\end{subfigure}
		\begin{subfigure}{0.48\linewidth}
			\centering
			\includegraphics[width=0.99\linewidth]{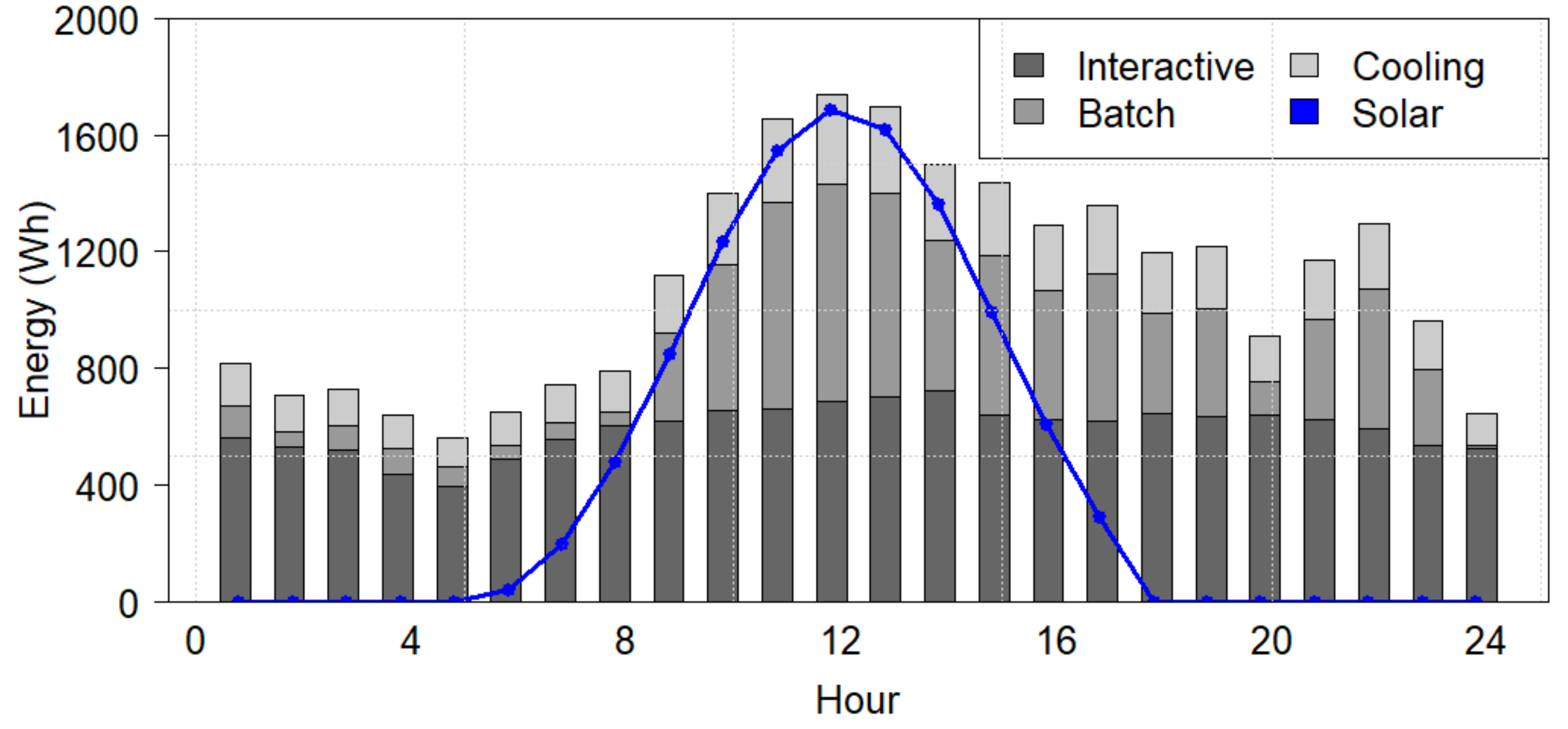}
			\caption{}
			\label{images/SA}
		\end{subfigure}
		\caption{\color{black}Results of (a) baseline HS (b) proposed approach GSA }
		\label{fig:SA}
	\end{figure*}

	\color{black} Figure \ref{images/noSA} shows the baseline energy consumption of interactive workloads, batch workloads and cooling during the observed time period (one day). The blue line shows the actual renewable power production. We consider the day in autumn time for Denver city. In this season, the day-length is about 12 hours, which is shorter than the summer time but longer than the winter time. In the investigated day, the system is consuming brown energy at \textit{night time} from hour 0:00 to hour 5:00 and hour 18:00 to hour 24:00. The solar energy is available at \textit{daytime} during hours 6:00 to 17:00. 
	Even with taking advantage of VM consolidation and host scaling, the solar energy consumption of the system is not fully utilized. For example, at hour 11:00, the total energy consumption of the system is about 1400 Wh, while the available solar energy is more than 1500 Wh. \color{black}
	%
	%
	%
	%
	%
	%
	%
	%
	%
	
	Figure \ref{images/SA}  demonstrates the energy consumption of GSA approach by using Algorithms 1 to 4. The blue line still shows the actual renewable power production, but the decision making is happening based on our SVM prediction model. We can observe that the power consumption of the batch workloads during 0:00 to 8:00 has been reduced, which results from the deferring operations: batch workloads are deferred to the time when solar energy is available, e.g. hour 6:00. Some batch workloads are still executed during hours from 0:00 to 8:00 due to the deadline constraints, which cannot be deferred to the time when renewable energy is available. Thus, we can find that the brown energy usage during 0:00 to 8:00 has been reduced compared in Figure \ref{images/noSA}. For example, at hour 1:00, the total power is reduced from 1221 to 815 Watts. 
	
	During the time when solar energy is available, GSA approach has improved the usage of renewable energy, in which the energy consumption follows the line the of predicted renewable energy. For instance, at hour 11:00, the usage of solar energy is increased from 1387 Wh to 1544 Wh compared with Figure \ref{images/noSA}. 
	
	We also note that the power consumption during the time when brown energy is the only source of power supply, the energy is also reduced, which exploits the brownout mechanism to reduce the energy consumption. For instance, the power at hour 18:00 is decreased from 1391 Watts to 1195 Watts.
	
	\color{black}Combining the results in Figure \ref{images/noSA} and \ref{images/SA}, we conclude that the GSA approach can improve the usage of renewable energy and reduce the usage of brown energy. To be noted, we choose the time in autumn for our experiments, as the renewable energy production in autumn is close to the time in spring, and these two seasons can represent renewable energy production about half a year. As for summer time, the day-length is extended, e.g. from hour 5:00 to hour 20:00, which means the workloads can have more possibilities to be deferred rather than being executed as soon as possible, and thus the utilization of renewable energy can be further improved. However, in the winter time, the day-length is reduced, e.g. only from hour 7:00 to 16:00. In this season, the utilization of renewable energy is not as good as in the autumn and summer seasons. \color{black}
	
	\begin{figure}[!ht]
		\centering
		\includegraphics[width=0.99\linewidth]{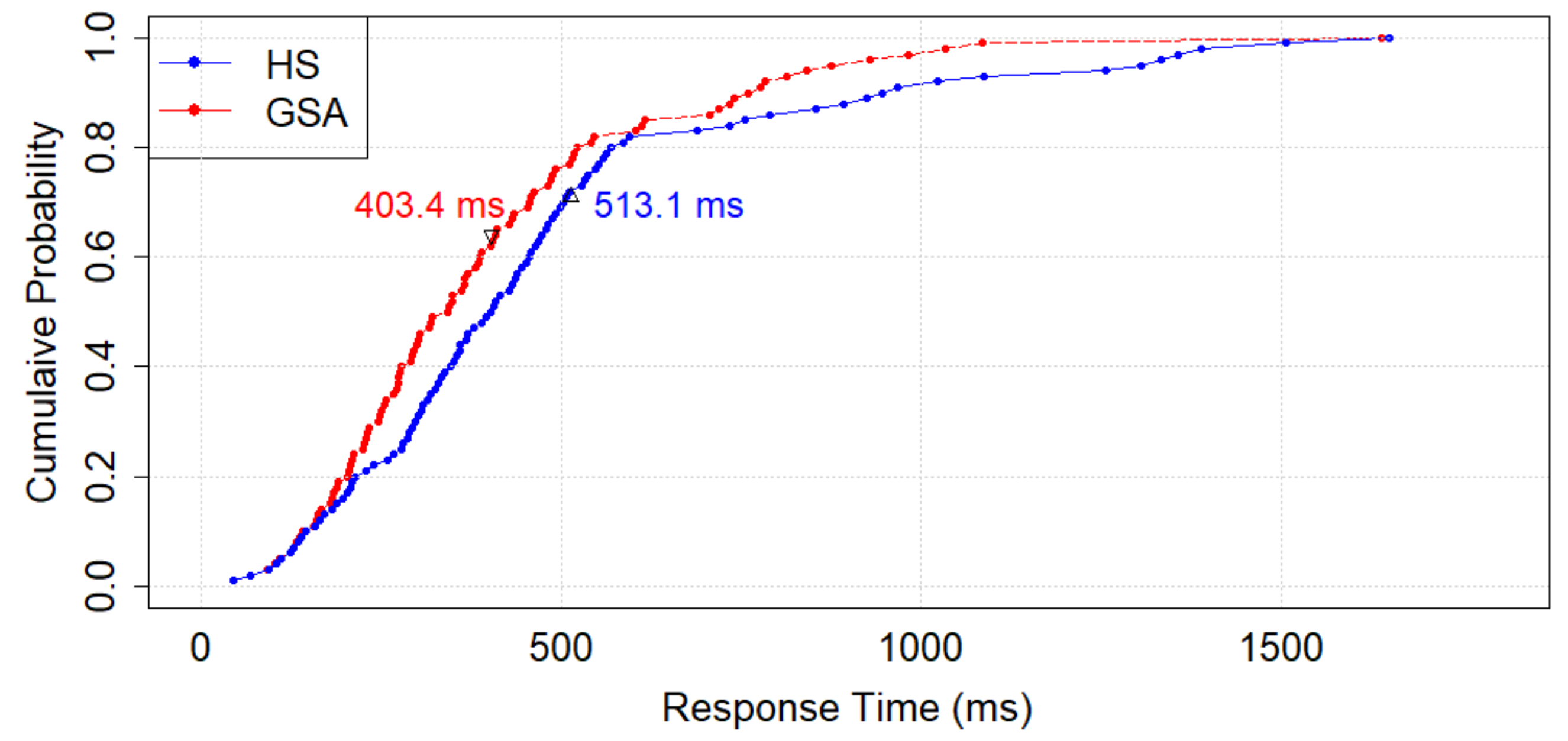}
		
		\caption[VarPerOptCom]{\color{black}Comparison of response time for interactive workloads\color{black}}

		\label{fig:responseTime}
	\end{figure}
	
	The average response time and \color{black}cumulative distribution function (CDF) \color{black}of response time for interactive workloads in our proposed approach and baseline are illustrated in Figure \ref{fig:responseTime}.
	The \color{black}average response time \color{black}of GSA is 403.4 ms, which is less than 80\% of the HS (513.1 ms)\color{black}. 
	\color{black}In GSA approach, the results also show that 95\% requests are responded in 900 ms, and 99\% requests are responded within 1000 ms second. While in the HS approach, only 91\% requests are responded within 1000 ms.
	It shows that GSA approach reduces brown energy usage while ensuring the QoS. The reason is that the brownout approach can relieve the overloaded situation, thus ensuring the response time. 
	
	\begin{figure}[!ht]
		\centering
		\includegraphics[width=0.99\linewidth]{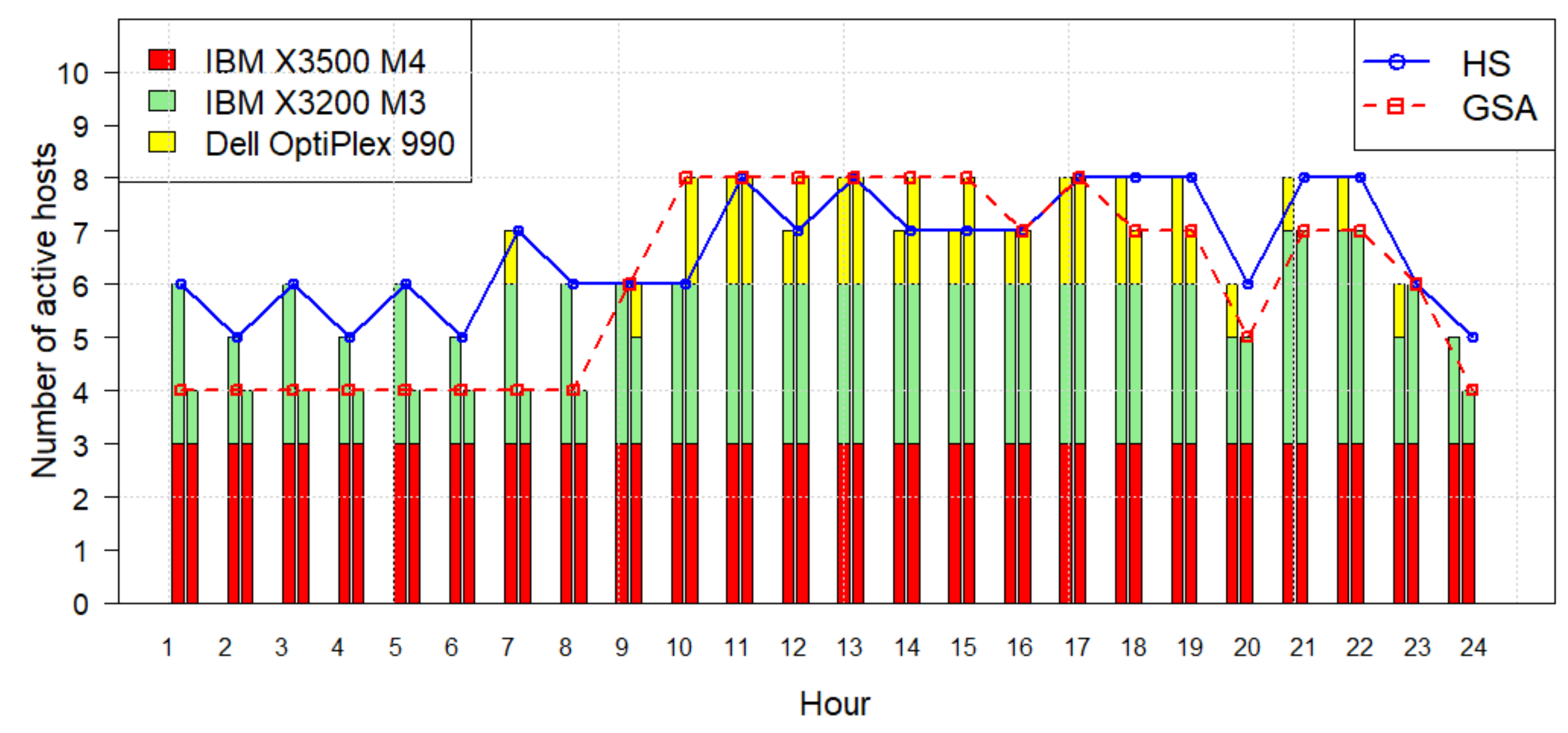}
		
		\caption[VarPerOptCom]{\color{black}Comparison of number of active hosts}

		\label{images/numberofactivehosts}
	\end{figure}

	\begin{figure*}[t!]
		\centering
		\begin{subfigure}{0.33\linewidth}
			\centering
			\includegraphics[width=0.99\linewidth]{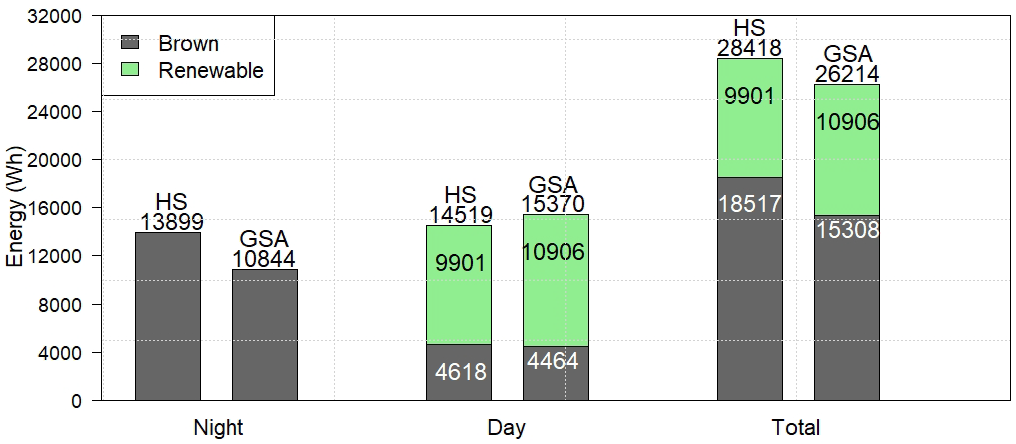}
			\caption{Autumn time in Denver city}
			\label{fig:energy}
		\end{subfigure}
		\begin{subfigure}{0.33\linewidth}
			\centering
			\includegraphics[width=0.99\linewidth]{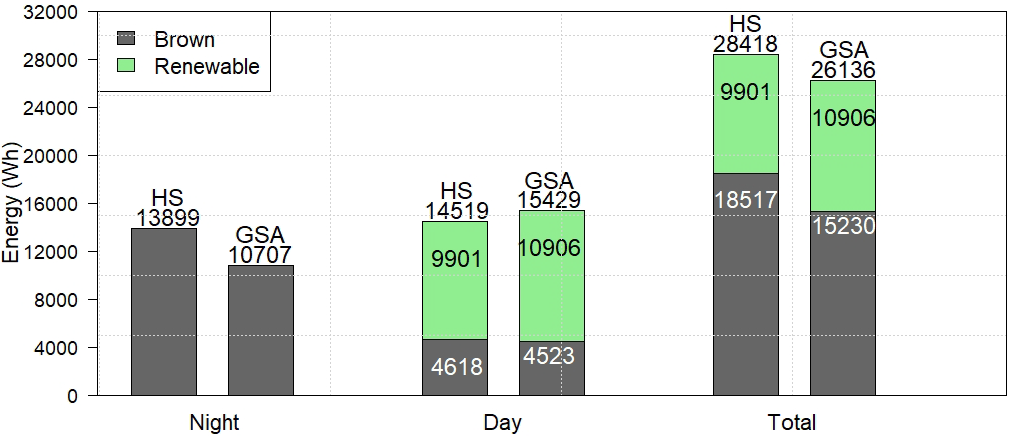}
			\caption{Longer deadline of batch workloads}
			\label{fig:energyddl}
		\end{subfigure}
		\begin{subfigure}{0.33\linewidth}
			\includegraphics[width=0.99\linewidth]{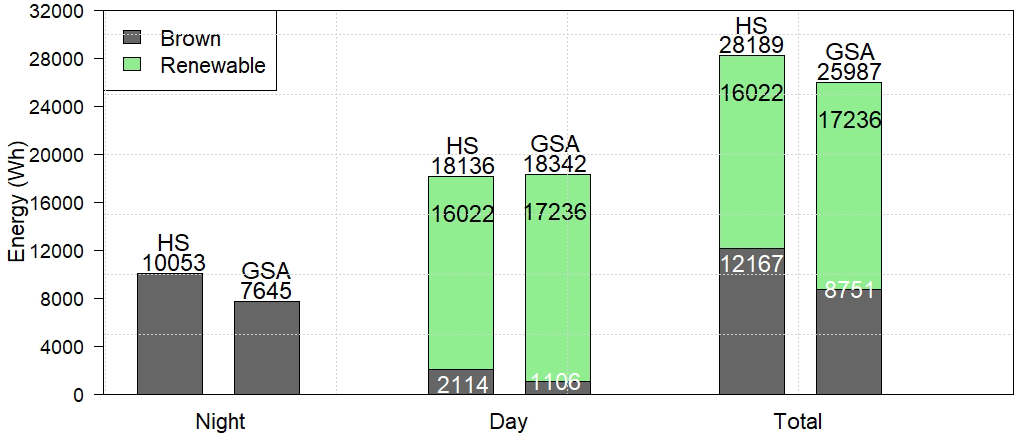}
			\caption{Longer daytime in summer}
			\label{fig:energydaytime}
		\end{subfigure}
		\vspace{-0.3cm}
		\caption{\color{black}Power usage comparison of different types of energy sources}
		\label{fig:eneryuse}
		\vspace{-0.5cm} 
	\end{figure*}
	
	\color{black}To illustrate the reason for power reduction by GSA approach, Figure \ref{images/numberofactivehosts} compares the active hosts during the observed time period between the HS and GSA approaches, as switching the idle hosts into the low power mode is the most effective way to save power. The results demonstrate that GSA approach uses fewer hosts during the time period when renewable energy is not sufficient, e.g. hours from 0:00 to 8:00 and 18:00 to 24:00, while when the renewable energy is available, more hosts are scaled in to utilize more renewable energy, such as the time from 10:00 to 15:00. In this way, the power of all the active hosts is reduced and the usage of renewable energy is improved. 
	Figure 7 also shows the type of active hosts in HS and GSA approaches. Based on the results, we can notice that the deactivation and activation are mainly operated on machines type of IBM X3200 M3 and Dell OptiPLex 990, which have less capacity than IBM X3500 M4 as shown in Table 2. The main reason is that these two types of hosts can host fewer VMs than IBM X3500 M4, and the VMs can be more easily consolidated to other machines. Therefore, after consolidation, the idle hosts can be switched to the low power mode.  \color{black}

	%
	%
	%
	
	Figure \ref{fig:energy} demonstrates the comparison of brown and renewable energy usage. During the night time (0:00 to 5:00 and 18:00 to 24:00), both approaches only use brown energy. Benefiting from proposed algorithms, our approach reduces the brown energy usage by 28\% from 13.9 kWh to 10.8 kWh.  During the daytime (6:00 to 17:00), both renewable energy and brown energy are used in two approaches. GSA approach consumes 5\% more total energy in the daytime, while it uses 10\% more renewable energy than the baseline from 9.9 KWh to 10.9 KWh. In total power usage comparison, the brown energy usage is reduced 21\%, and the renewable energy usage is improved 10\%.

	\color{black}To further investigate the impacts on different configurations, we conduct another two experiments by changing the deadline of batch workloads and the availability of renewable energy. Due to the page limitation, we only demonstrate the results of different types of power usage like in Figure \ref{fig:energy}.
	
	\textit{Longer deadlines of batch workloads:} we generate the deadline for processing jobs as uniform distribution with $\mu=7$ hours and $\sigma = 2$ hours in $N(\mu, \sigma^2)$, which has a longer deadline than the previous configurations. The other configurations are the same as the previous experiments. In Figure \ref{fig:energyddl}, compared with HS, we can notice that the brown energy usage in the night time is reduced from about 13.9 kWh to 10.7 kWh by GSA, renewable energy usage is increased from 9.9 kWh to 10.9 kWh, and total energy usage is reduced about 8\% by GSA. Compared with the results in Figure \ref{fig:energy} where the batch workloads have shorter deadlines, the brown energy consumption has been reduced, e.g. from 10.8 kWh to 10.7 kWh in the night time, which shows that longer deadline is helpful to reduce brown energy usage. \color{black}The reason is that more batch workloads are deferred to the daytime when the renewable energy is available. Therefore, the brown energy usage in the night time is reduced. However, the brown energy usage in the daytime is increased a bit while the total brown energy usage is reduced.\color{black}

	%
	%
	%
	\color{black}
	\textit{Longer daytime in summer with more varied solar power:} since the solar power dataset of Denver city has sufficient data, thus we still use the dataset, but we change the season from autumn to summer, which has longer daytime that starts from hour 5:00 and ends at hour 19:00 and can represent solar power with variability. The other settings are configured the same as settings in Figure \ref{fig:energy}. Figure \ref{fig:energydaytime} shows the power usage comparison with longer daytime. It can be observed that GSA can reduce the total brown energy usage from 12.2 kWh to 8.8 kWh and renewable energy usage can be improved from 16.0 kWh to 17.2 kWh. Compared with the results in Figure \ref{fig:energy} and \ref{fig:energyddl}, the renewable energy percentage of total energy usage has been significantly increased, as the daytime is extended and the amount renewable energy is more sufficient in the summer time compared with the autumn time.
	
	\color{black}To evaluate the brownout impacts on microservices, we use the average deactivation percentage (the average number of deactivated microservices divided by the total number of microservices during the observation time). The higher the average deactivation percentage, the more microservices are deactivated, and vice verse.
	 The average deactivation percentage for cases in Figure \ref{fig:energy} and Figure \ref{fig:energyddl} are  11.2\% and 10.8\%, respectively, and in the case of Figure \ref{fig:energydaytime}, the value is 10.1\%. Based on the results, we can observe that the longer deadline of batch workloads and the longer daytime result in a lower number of deactivated microservices. The reason is that  a higher renewable energy availability can support more active microservices.  \color{black}
 	
	%
	%
	%
	
	\color{black}In summary, experiments show that GSA approach  can improve the renewable energy usage for both interactive and batch workloads by applying brownout mechanism and deferring the execution of batch workloads.  
	Our proposed approach can switch more machines into low power mode when renewable energy is not sufficient while the QoS of workloads is also ensured. 
	
	\section{Conclusions and Future Work} \label{sec:conclusion}
	Our self-adaptive approach for managing applications and harnessing renewable energy brings up many opportunities to optimize the energy efficiency problem in cloud computing environment. In this paper, we proposed a multiple layers perspective model for interactive and batch workloads by considering renewable energy. 
	We also introduced a self-adaptive and renewable energy-aware approach deriving from the perspective model. The proposed approach improves the usage of renewable energy and reduces the usage of brown energy while ensuring the QoS requirement of workloads. \color{black}We apply a solar radiation prediction method to predict solar power at Denver City and integrate it into our proposed approach. \color{black}We utilize brownout mechanism to dynamically deactivate/activate optional components in the system for interactive workloads and use a deferring algorithm to defer the execution of batch workloads to the time when renewable energy is available. VM consolidation and host scaling are also applied to reduce the number of active hosts.  
	
	We developed a prototype system to evaluate the performance of our proposed approach. In the prototype system, the physical resources are managed by OpenStack and the services are managed by Docker Swarm. We take advantage of the APIs from these platforms to monitor, manage, and provision the resources to services. The effectiveness of our proposed approach is showed through the experimental evaluations with a microservices-based web system and the workloads from real traces. The results show that our proposed approach is able to improve the usage of renewable energy while satisfying the constraints of workloads. 
	
	
	
	\color{black}As future work, we would like to include the battery model \color{black}in \cite{goiri2013parasol}, \color{black}which can store renewable energy and improve energy usage. We also plan to extend our prototype system for multiple clouds in the different time zones to support workload shifts in data centers and minimize the carbon footprint in a global view.
	
	Fog and Edge computing extend the cloud services to the edge of the network, which can improve the user experience and system performance by reducing latency. However, the IoT and edge devices have power constraints, for example, they are powered by battery or they need to harness renewable energy. Therefore, the energy should be used in an efficient manner. The brownout approach can support to optimize the energy usage for these devices by temporarily deactivating some optional application components. As another future work, we would like to apply the brownout approach to mobile edge computing for managing the energy usage of IoT or edge devices.  \color{black}
	

	
	%

	\appendices
	

	\section*{Acknowledgments}
	This work is supported by Key-Area Research and Development Program of Guangdong Province (NO. 2020B010164003), Science and Technology Development Fund of Macao S.A.R (FDCT) under number 0015/2019/AKP, Shenzhen Discipline Construction Project for Urban Computing and Data Intelligence, SIAT Innovation Program for Excellent Young Researchers and ARC Discovery Project. 

	\ifCLASSOPTIONcaptionsoff
	\newpage
	\fi

	
	
	%
	%
	%
	
	\bibliographystyle{IEEEtran}
	\bibliography{library}
	
	%

	
	

	\begin{IEEEbiography}
		[{\includegraphics[width=1in,height=1.25in,clip,keepaspectratio]{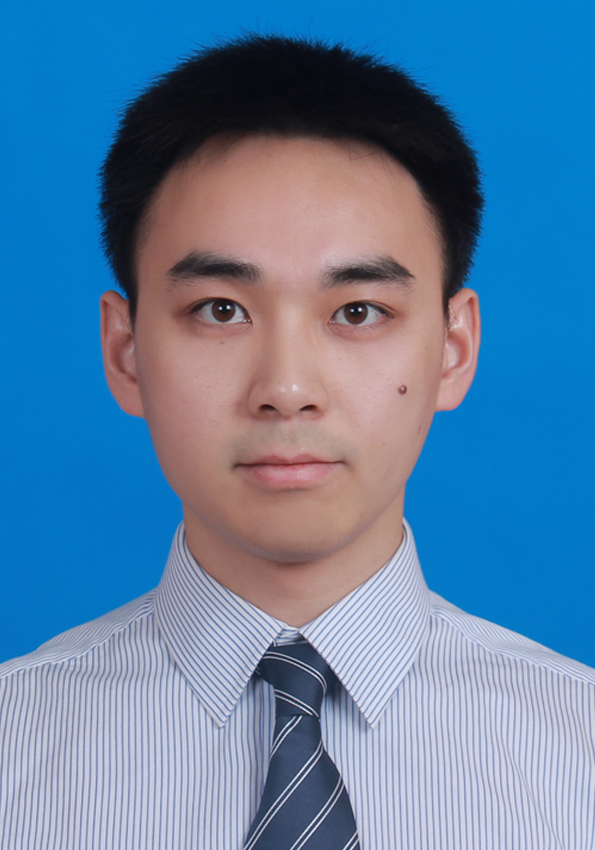}}]{Minxian Xu} is currently an assistant professor at Shenzhen Institutes of Advanced Technology, Chinese Academy of Sciences. He received the BSc degree	in 2012 and the MSc degree in 2015, both in software engineering from University of Electronic Science and Technology of China. He obtained his PhD degree from the University of Melbourne in 2019. His research interests include resource scheduling and optimization in cloud computing. He has co-authored 20+ peer-reviewed papers published in prominent international journals and conferences, such as ACM Computing Surveys, IEEE Transactions on Sustainable Computing, IEEE Transactions on Automation Science and Engineering, Journal of Parallel and Distributed Computing, Concurrency and Computation: Practice and Experience, International Conference on Service-Oriented Computing. His Ph.D. Thesis was awarded the 2019 IEEE TCSC Outstanding Ph.D. Dissertation Award. More information can be found at: minxianxu.info.
	\end{IEEEbiography}
	\begin{IEEEbiography}
		[{\includegraphics[width=1in,height=1.25in,clip,keepaspectratio]{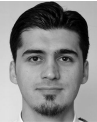}}]{Adel Nadjaran Toosi}
		has joined Faculty of Information Technology at Monash University as a lecturer in May 2018. Before he joined Monash University, he worked as a Research Fellow in the Cloud Computing and Distributed Systems (CLOUDS) Laboratory in the School of Computing and Information Systems (CIS) at the University of Melbourne for more than three years. He received his Ph.D. degree in Computer Science and Software Engineering from the University of Melbourne in 2015. His thesis was one of the two theses nominated for the Chancellor's Prize for Excellence in the Ph.D. Thesis and John Melvin Memorial Scholarship for the Best Ph.D. Thesis in Engineering. Adel has made significant contributions to the areas of resource management and software systems for cloud computing. His research interests include Distributed Systems, Cloud Computing, Software-Defined Networking (SDN), Green Computing, and Soft Computing. He is currently working on resource management for Software-Defined Networking (SDN)-enabled cloud computing environments. For more details, please visit his homepage: http://adelnadjarantoosi.info.
		
	\end{IEEEbiography}
	\begin{IEEEbiography}
		[{\includegraphics[width=1in,height=1.25in,clip,keepaspectratio]{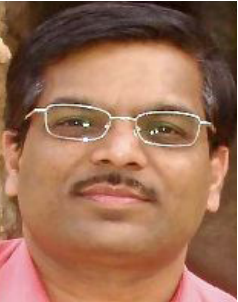}}]{Rajkumar Buyya}
		is a Redmond Barry Distinguished Professor and Director of the Cloud Computing and Distributed Systems (CLOUDS) Laboratory at the University of Melbourne, Australia. He is also serving as the founding CEO of Manjrasoft, a spin-off company of the University, commercializing its innovations in Cloud Computing. He served as a Future Fellow of the Australian Research Council during 2012-2016. He has authored over 625 publications and seven text books including "Mastering Cloud Computing"  published by McGraw Hill, China Machine Press, and Morgan Kaufmann for  Indian, Chinese and international markets respectively.  He is one of the highly cited authors in computer
		science and software engineering worldwide  (h-index=136, g-index=300, 98,800+ citations). Dr. Buyya is
		recognized as a ”Web of Science Highly Cited Researcher” for four
		consecutive years since 2016, a Fellow of IEEE, and Scopus Researcher of the Year 2017 with Excellence in Innovative Research Award by Elsevier for his outstanding contributions to Cloud computing. He served as the founding Editor-in-Chief of the IEEE Transactions on Cloud Computing. He is currently serving as Co-Editor-in-Chief of Journal of Software: Practice and Experience, which was established ~ 50 years ago.
		For further information on Dr. Buyya, please visit his cyberhome:
		www.buyya.com
		
	\end{IEEEbiography}
	
\end{document}